\documentstyle[prb,aps,epsfig,preprint]{revtex}
  
\newcommand{\be}{\begin{equation}}
\newcommand{\ee}{\end{equation}}

\newcommand{\ba}{\begin{eqnarray}}
\newcommand{\ea}{\end{eqnarray}}

\def\sg{{\sigma}}
\def\ros{{\rho(\sg)}}
\def\br{{\mathbf{r}}}
\def\fid{{f_{id}}}
\def\fex{{f_{ex}}}
\def\delro{{\delta\rho}}
\def\tro{{(T,[\rho])}}

\begin{document}
\bibliographystyle{unsrt}

\title{Phase diagrams of polydisperse van der Waals fluids}

     \author{L. Bellier-Castella$^{\dag}$, H. Xu $^{\dag}$ and M. Baus
$^{\dag\dag}$}

\maketitle

\noindent
\begin{center}
$^{\dag}$ D\'epartement de Physique des Mat\'eriaux (UMR 5586 du CNRS),\\
Universit\'e Claude Bernard-Lyon1, 69622 Villeurbanne Cedex, France\\
$^{\dag\dag}$ Physique des Polym\`eres, Universit\'e Libre de Bruxelles,\\
Campus Plaine, CP 223, B-1050 Brussels, Belgium\\
\end{center}
\vspace{2truecm}

\begin{abstract}
  The phase behavior of a system composed of spherical particles with a
monomodal size distribution is investigated theoretically within the
context of the van der Waals approximation for polydisperse fluids.
It is shown how the binodals, spinodals, cloud-point and shadow curves
as well as all the (polydispersity induced) critical points can be
obtained for a variety of interaction potentials. The polydispersity
induced modifications of the phase diagram (even for a polydispersity
index $I$ as small as $I\approx 1.01$) should be observable in some
colloidal dispersions.  
\end{abstract}

\pagebreak
\section{Introduction}
\label{sec1}
\vspace{1truecm}
   Many of the systems encountered in soft-matter physics (such as e.g.
colloidal dispersions~\cite{russ},
liquid crystals~\cite{chan}, polymeric melts~\cite{elia})
are fluids exhibiting one or several ``polydispersities" because,
by the very nature of their production process, these fluids are 
collections of complex molecular objects which
are not strictly identical one to another.
When these objects can be grouped into $n$ sets of identical objects
they form an ordinary $n$-component mixture~\cite{deho}
whereas here we will focus instead
on the case where the differences between these objects (e.g. differences
in size, shape, surface charge, chemical composition, etc.) are distributed
in an almost continuous manner in which case they can be described as
``continuous'' mixtures containing infinitely many components~\cite{aris}.
In the present
study we will consider one of the simplest situations where the
system is composed of spherical molecular objects which differ only in size.
This excludes the liquid crystals and polymeric melts from our consideration.
We will hence focus our attention on colloidal dispersions of spherical
particles with a continuous size distribution~\cite{hunt}. 
In practice, this size
distribution can be either monomodal (i.e. having only one maximum)
or multimodal (i.e. having several maxima). We will again consider
only the simplest case of a system with a monomodal size distribution~\cite{hunt}.
Such systems can hence be considered as polydisperse generalizations
of single component systems (whereas systems with a multimodal
size distribution are generalizations of multi-component mixtures).
When this monomodal size distribution is strongly peaked around its maximum
one often refers to the corresponding colloidal dispersion as being
monodisperse~\cite{ref7}. At present, such monodisperse colloidal dispersions
are often used as testing ground for the study of various aspects
of liquid state theory~\cite{ref7}. Indeed, by clever chemical engineering the colloidal
particles can often be endowed with properties which are outside
the range of ordinary single component systems (e.g. they can be
prepared such as to have hard-sphere like repulsions and short-range
 attractions~\cite{ref7}). Nevertheless, even in those colloidal dispersions
which are usually denoted as monodisperse there always remains
a residual size distribution. While the influence of this (unavoidable)
polydispersity on the behavior of a single phase is well understood~\cite{lero},
the first-principles study of its influence on a phase separation
process is on the contrary still under active scrutiny~\cite{bart}. Because
the theoretical study of phase separations in polydisperse systems is
faced with rather heavy technical problems we will consider here only
the simplest possible phase separation, namely a fluid- fluid phase transition~\cite{deho}.
The theoretical tool used to describe this transition will again be the
simplest one, namely the van der Waals (vdW) approximation~\cite{lovet}. It is indeed well
known that the vdW-approximation provides us with a description of the
phase separation of a fluid into a dilute fluid (or gaz phase) and a dense fluid
(or liquid phase) which is both simple and physically sound. It moreover
involves a (mean-field) critical point which constitutes an important
feature of many phase diagrams~\cite{deho}. The vdW-approximation is also very
robust and can be generalized in several ways, including for polydisperse
systems~\cite{gual,xu}. In the present study we will therefore focus our attention on the
modifications brought about by the polydispersity to the binodal and
critical point of a vdW-fluid with a monomodal size distribution.
Such a description will of course not always be fully realistic for
a given colloidal dispersion but we may hope that it retains the qualitative
correctness which has characterized the widespread use of vdW-like
descriptions in the past~\cite{lovet}. Since the present vdW-approximation is restricted
to fluid phases only it will not be possible here to sort out which parts
of the phase diagrams obtained are metastable with respect to the solid
phases. By focusing on the fluid phases only, we are nevertheless able to
show how the many technical problems raised by the study of phase equilibria
in polydisperse systems can be solved, paving hereby the way for their later
extensions to other phases and more involved descriptions.

In section~\ref{sec2} we summarize the thermodynamic conditions of phase
coexistence in a polydisperse system in terms of the system's free-energy~\cite{deho,aris}
which has here to be viewed as a functional of the density distribution among
the species. We also deduce the thermodynamic stability conditions
of a polydisperse system from the convexity of this free-energy functional.
This procedure, which is intrinsic in the sense that it does not rely on
taking the limit of an infinitely many component system is, as far as
we are aware, new. In section~\ref{sec3} we introduce the vdW free-energy
in a form already used elsewhere~\cite{xu} (for the study of weakly polydisperse
systems) except that here we include the cases where the polydispersity
also affects the amplitude of the interaction potential. The monomodal
parent phase size distributions considered in this work are introduced
in section~\ref{sec4} while section~\ref{sec5} is devoted to the solution
method to be used subsequently for obtaining the numerical solutions
of the integral equations which, for a polydisperse system, replace the algebraic equations governing the phase equilibria. This method
can be generalized to any system for which the free-energy has an excess
(over ideal) part which depends on a finite number of (generalized) moments
of the polydispersity distribution. In section~\ref{sec6} we show how
the thermodynamic stability conditions of section~\ref{sec2} can be
reduced to closed algebraic equations whenever the excess free-energy
is of this particular type. In section~\ref{sec7} we present explicit
phase diagrams for interaction potentials with and without amplitude
polydispersity. The critical behavior of a polydisperse vdW fluid
is studied further in section~\ref{sec8} while the final section~\ref{sec9}
contains our conclusions.\par

\section{Phase equilibrium in a polydisperse fluid}
\label{sec2}
The statistical mechanical description of a polydisperse equilibrium
system~\cite{xu} is equivalent to a density functional theory for a
system whose number density, say $\rho(\br,\sg)$, depends besides the
position variable $\br$ (assuming spherical particles) moreover on
a continuous species label $\sg$ (considered to be dimensionless).
Such a theory is completely determined by the knowledge of its
intrinsic Helmholtz free-energy ($F$) per unit volume ($V$),
$f=F/V$, which is a function of the temperature ($T$)
and a functional of the
number density, $f=f(T,[\rho])$, where $[\rho]$ indicates a
functional dependence on $\rho(\br,\sg)$.
All thermodynamic properties can then be deduced from 
$f(T,[\rho])$ by functional differentiation~\cite{baus}. Below we will consider
only (uniform and isotropic) fluid phases for which $\rho(\br,\sg)$
is independent of $\br$, hence $\rho(\br,\sg)\equiv\ros$. 

\subsection{The thermodynamic functionals}
\label{sec2p1}
The phase behavior of a polydisperse fluid is determined by the chemical
potential of the various species $\sg$, $\mu=\mu(\sg,T,[\rho])$, and the
pressure, $p=p(T,[\rho])$, of the system. These functionals can be obtained from
$f(T,[\rho])$ by the usual rules~\cite{baus}, suitably extended to a continuous
mixture~\cite{aris},
as:
\be
\mu(\sg,T,[\rho])=\delta f(T,[\rho])/\delta \ros\ ;
\label{eq21}
\ee 
\be
p(T,[\rho])=\int d\sg\ros\mu(\sg,T,[\rho])-f(T,[\rho])\ .
\label{eq22}
\ee
In what follows it will be convenient to split these quantities into an
explicitly known ideal gaz part~($id$) and a generally unknown excess
part~($ex$), for ex. $f=\fid+\fex$, with:
\be
\fid(T,[\rho])=k_B\,T\,\int d\sg\ros\left\{\ln\left(\Lambda^3(\sg)\ros\right)
\;-1\right\}
\label{eq23}
\ee
\be
\mu_{id}(\sg,T,[\rho])=k_B\,T\,\ln\left\{\Lambda^3(\sg)\ros\right\}
\label{eq24}
\ee
\be
p_{id}(T,[\rho])=k_B\,T\,\int d\sg\ros
\label{eq25}
\ee
where $k_B$ denotes Boltzmann's constant, $\Lambda(\sg)$ the thermal
de Broglie wavelength of species $\sg$, and the integrals over $\sg$ extend
over the whole domain of definition of $\ros$.

\subsection{Phase coexistence conditions}
\label{sec2p2}
When for a given $T$, a parent phase of density $\rho_0(\sg)$,
phase separates into $n$ daughter phases of density $\rho_i(\sg)$
($i=1,\ldots,n$), the thermodynamic conditions of phase equilibrium~\cite{deho}
imply the equality of the pressures:
\be
p(T,[\rho_1])=p(T,[\rho_2])=\ldots=p(T,[\rho_n])
\label{eq26}
\ee
and of the chemical potentials:
\be
\mu(\sg,T,[\rho_1])=\mu(\sg,T,[\rho_2])=\ldots=\mu(\sg,T,[\rho_n])
\label{eq27}
\ee
for each species $\sg$. It will be convenient to rewrite,
$\rho_j(\sg)=\rho_j h_j(\sg)$, where $\rho_j=\int d\sg\rho_j(\sg)$ is
the average density and $h_j(\sg)$ the normalized (viz.
$\int d\sg h_j(\sg)=1$)
polydispersity distribution of phase $j$ ($j=0,1,2,\ldots,n$).
The phase separation is moreover constrained by the conservation of
the total number of particles of each species $\sg$:
\be
h_0(\sg)=\sum_{i=1}^{n} x_i h_i(\sg)
\label{eq28}
\ee
where, $x_i=N_i/N_0$, is the ratio of the total number of particles
($N_i$) in phase $i$ to the total number of particles in the parent
phase ($N_0$). Moreover, the conservation of the total volume occupied
by the parent phase can be written:
\be
v_0=\sum_{i=1}^{n} x_i v_i
\label{eq29}
\ee
where $v_j=1/\rho_j$ ($j=0,1,2,\ldots,n$). Finally, the normalization of the
$h_j(\sg)$ in~(\ref{eq28}) implies:
\be
1=\sum_{i=1}^{n} x_i
\label{eq210}
\ee
which expresses the conservation of the total number of particles.

In principle, given $T$, $\rho_0$ and $h_0(\sg)$, one has to solve the system
of equations (\ref{eq26}-\ref{eq210}) for the $\rho_i$ and $h_i(\sg)$
($i=1,2,\ldots,n$). Even when starting from a relatively simple $f(T,[\rho])$
this turns out to be a rather formidable task because (\ref{eq26}-\ref{eq27})
are no longer algebraic equations (as would be the case for ordinary mixtures)
but become here integral equations for the $h_i(\sg)$. In order
to simplify the situation somewhat we will restrict ourselves here to two-phase
coexistences only, i.e. to $n=2$. In a polydisperse system $n$ can in principle
be arbitrarily large because the Gibbs phase rule~\cite{deho} does not restrict the value of
$n$ in a system of infinitely many species~\cite{aris}. In practice, however, a polydisperse
system does rarely use this infinite number of thermodynamic degrees of freedom.
Since these multiple-phase coexistences are expected to occur at low temperatures,
the restriction to two-phase coexistences ($n=2$) implies that the value of $T$
should be chosen high enough (see below). Note that this is consistent with
our earlier restriction (see above) to fluid phases because for low temperatures
some of the (multiple) fluid phases may have to compete with some of the solid phases.

For a two-phase coexistence ($n=2$) the number fractions $x_i$ ($i=1,2$) can
be obtained from~(\ref{eq29}-\ref{eq210}) as:
\be
x_1=\frac{v_0-v_2}{v_1-v_2},\ x_2=\frac{v_0-v_1}{v_2-v_1} 
\label{eq211}
\ee
expressing the lever rule~\cite{deho}, whereas $h_2(\sg)$ can be eliminated by using
moreover~(\ref{eq28}):
\be
h_2(\sg)=\frac{v_2-v_1}{v_0-v_1}\ h_0(\sg)+\frac{v_2-v_0}{v_1-v_0}\ h_1(\sg)
\label{eq212}
\ee
where $v_0=1/\rho_0$ and $h_0(\sg)$ are given parent phase data. To find
$v_1=1/\rho_1$, $v_2=1/\rho_2$ and $h_1(\sg)$ we need three relations.
To this end we use~(\ref{eq24}) to rewrite~(\ref{eq27}) for $n=2$ as:
\be
\rho_1(\sg)=\rho_2(\sg)\exp\left\{\beta\Delta\mu_{ex}(\sg,T,[\rho_1,\rho_2])
\right\}
\label{eq213}
\ee
where $\beta=1/k_B\,T$ and : 
\be
\Delta\mu_{ex}(\sg,T,[\rho_1,\rho_2])=\mu_{ex}(\sg,T,[\rho_2])
-\mu_{ex}(\sg,T,[\rho_1])
\label{eq214}
\ee
with $\mu_{ex}$ being the excess part of $\mu(\sg,T,[\rho])$. In terms
of the $h_i(\sg)$, eq.~(\ref{eq213}) becomes:
\be
h_1(\sg)=h_2(\sg)\;A_{12}(\sg,T;\rho_1,\rho_2;[h_1],[h_2])
\label{eq215}
\ee
where $A_{12}$ is a shorthand notation for $A_{12}=(v_1/v_2)\exp\{\beta
\Delta\mu_{ex}\}$. Eliminating $h_2(\sg)$ from~(\ref{eq215}) by using~(\ref{eq212})
one obtains finally:
\be
h_1(\sg)=h_0(\sg)\,H(\sg,T;\rho_0,\rho_1,\rho_2;[h_1],[h_0])
\label{eq216}
\ee
where, $H=(v_2-v_1)\,A_{12}/\{(v_0-v_1)+(v_2-v_0)A_{12}\}$, and it is understood
that $h_2(\sg)$ has also been eliminated from $A_{12}$ by using~(\ref{eq212}).
Eq.~(\ref{eq216}) is our first relation between $\rho_1$, $\rho_2$ and $h_1(\sg)$.
Given $T$, $\rho_0$ and $h_0(\sg)$, eq.~(\ref{eq216}) can in principle be solved
with respect to $h_1(\sg)$ for given $\rho_1$ and $\rho_2$ values. A first
relation between $\rho_1$ and $\rho_2$ can then be found by solving moreover:
\be
1=\int\,d\sg\,h_0(\sg)\,H(\sg,T;\rho_0,\rho_1,\rho_2;[h_1],[h_0])
\label{eq217}
\ee
which follows from~(\ref{eq216}) and the normalization of $h_1(\sg)$. Finally,
the system of equations is closed by solving moreover~(\ref{eq26}):
\be
p(T,[\rho_1])=p(T,[\rho_2])
\label{eq218}
\ee
expressing the equality of the pressures. It is obvious that solving a
(two-phase) coexistence problem for a polydisperse fluid is much more
involved than the corresponding problem for a monodisperse (single species)
fluid. This is due mainly to the central integral equation~(\ref{eq216})
which governs the change in the polydispersity or species distribution
between the parent phase and the two coexisting phases (cf~(\ref{eq212})),
a process usually called fractionation~\cite{bart}. The extension of
(\ref{eq216}-\ref{eq218}) for the case of a $n$-phase coexistence is straightforward.

\subsection{Thermodynamic stability conditions}
\label{sec2p3}
Of course, not for every prescribed \{$T,\rho_0(\sg)$\} will the corresponding
 phases exist or be thermodynamically stable. For this to be
the case, the free-energy density, $f(T,[\rho])$, must
remain a convex functional of $\ros=\rho_0(\sg)$, i.e. it must satisfy:
\be
f(T,[\rho+\lambda\delro])<\;\lambda f(T,[\rho+\delro])
+(1-\lambda)f(T,[\rho])
\label{eq219}
\ee
for any, $0<\lambda<1$, and for any change, $\delta\ros\neq 0$, of the
functional form of $\ros$. If we consider only infinitesimal changes,
$\delta\ros$, eq.~(\ref{eq219}) is equivalent to:
\be
0<\sum_{k=2}^{\infty}\frac{1}{k!}(\lambda-\lambda^k)\ \delta^k f(T,[\rho])
\label{eq220}
\ee
where $\delta^k f$ is the $k$-th functional variation of $f$:
\be
\delta^k f=\int d\sg_1...\int d\sg_k\; K_k(\sg_1,...\sg_k; T,[\rho])
\;\delta\rho(\sg_1)...\delta\rho(\sg_k)
\label{eq221}
\ee
and $K_k$ is an integral operator whose kernel is :
\be
K_k(\sg_1,...\sg_k; T,[\rho])=\frac{\delta^k f(T,[\rho])}
{\delta\rho(\sg_1)...\delta\rho(\sg_k)}\;\;.
\label{eq222}
\ee 
Hence, the condition for stability with respect to infinitesimal changes
($\ros\rightarrow\ros+\delro(\sg)$) reduces ($\delta^{k+1}f\ll\delta^k f$) to:
\be
0<\delta^2 f(T,[\rho]).
\label{eq223}
\ee
In a case of marginal stability there must exist at least one ``fluctuation"
$\delro(\sg)\ne 0$, say $\delro_0(\sg)$, such that $\delta^2 f(T,[\rho])=0$,
or explicitly:
\be
\int d\sg_1\int d\sg_2\; K_2(\sg_1,\sg_2; T,[\rho])
\;\delta\rho_0(\sg_1)\delta\rho_0(\sg_2)=0
\label{eq224}
\ee
for a given $T$ and $\ros$. Any solution $\delro_0(\sg)$ of~(\ref{eq224}) will
be called a critical fluctuation. For the system to remain stable with respect
to these critical fluctuations $\delro_0(\sg)$ we must have according
to~(\ref{eq220}), $\delta^3f=0$ (because the sign of $\delta^3f$ changes
with the sign of $\delro_0(\sg)$ while~(\ref{eq224}) does not fix the sign of $\delro_0(\sg)$ ) and $\delta^4 f>0$:
\be
\int d\sg_1\int d\sg_2\int d\sg_3\; K_3(\sg_1,\sg_2,\sg_3; T,[\rho])
\;\delta\rho_0(\sg_1)\delta\rho_0(\sg_2)\delta\rho_0(\sg_3)=0;
\label{eq225}
\ee
\ba
\int\ d\sg_1\ \int\ d\sg_2\ \int\ d\sg_3\ \int\ d\sg_4
\ K_4(\sg_1,\sg_2,\sg_3,
\sg_4; T,[\rho])
\nonumber\\
.\delta\rho_0(\sg_1)\ \delta\rho_0(\sg_2)\ \delta\rho_0(\sg_3)\ \delta\rho_0(\sg_4)\ >\ 0.
\label{eq226}
\ea
Note that in the present context the stability conditions~(\ref{eq223}-\ref{eq226})
imply stability with respect to changes in both the average density ($\delro(\sg)
=\delro\,h(\sg)$) and the composition ($\delro(\sg)=\rho\,\delta h(\sg)$).
The values of $T$ and $\ros$ for which~(\ref{eq224}) has a solution define a (generalized or polydisperse) spinodal whereas those values for which~(\ref{eq224}) and~(\ref{eq225}-\ref{eq226})
are simultaneously satisfied correspond to the critical states of the polydisperse
fluid. For a monodisperse fluid these conditions reduce to well known results
but for a polydisperse fluid they are much less well known and also much more
difficult to study. For instance, for eq.~(\ref{eq224}) to have a solution,
the Fredholm determinant~\cite{gour} of the integral operator $K_2$ must vanish. This
functional determinant however is equivalent to an infinite series of
ordinary determinants~\cite{gour}. As shown by Kincaid et al.~\cite{kinc},
in order to draw any
conclusion from such a series of determinants one must have at its disposal
a smallness parameter allowing one to truncate this series. For the weakly
polydisperse fluids considered by Kincaid et al. ~\cite{kinc} there is such a parameter
but for the strongly polydisperse systems to be considered here there is none.
Needless to say that in order to make progress we will need further
simplifying assumptions. To this end we will introduce in the next section
a model expression for $f(T,[\rho])$ which is simple enough to allow us
to tackle the various technical problems encountered above in the study
of phase equilibria in polydisperse fluids.

\section{The polydisperse van der Waals fluid}
\label{sec3}
An explicit expression for $f\tro$, which is both relatively simple and physically
sound, can be obtained from the van der Waals (vdW) theory~\cite{lovet}. For a polydisperse
fluid the vdW free-energy reads~\cite{gual,xu}:
\ba
f(T,[\rho])=k_B\,T\int d\sg\ros\{\ln(
\frac{\Lambda^3(\sg)\ros}{E[\rho]})-1\}\nonumber\\
+\frac{1}{2}\int d\sg\int d\sg'\,V(\sg,\sg')\ros\rho(\sg')
\label{eq31}
\ea
where the hard repulsions between the particles are taken into account via
the usual vdW excluded volume correction~\cite{lovet}, $E[\rho]$:
\be 
E[\rho]=1-\int d\sg\,v(\sg)\ros
\label{eq32}
\ee
$v(\sg)$ being the volume of a particle of species $\sg$,
while the cohesion energy~\cite{lovet} resulting from the interparticle attractions
described by $V_A(r;\sg,\sg')$ is given in the vdW mean field approximation
by the second term of~(\ref{eq31}) with:
\be
V(\sg,\sg')=\int d\br\,V_A(r;\sg,\sg').
\label{eq33}
\ee
From~(\ref{eq31}) and~(\ref{eq21}) we obtain:
\ba
\mu(\sg,T,[\rho])=k_B\,T\ln\{
\frac{\Lambda^3(\sg)\ros}{E[\rho]}\}
+k_B\,T\frac{v(\sg)}{E[\rho]}\int d\sg'\rho(\sg')\nonumber\\
+\int d\sg'\,V(\sg,\sg')\rho(\sg')
\label{eq34}
\ea
while~(\ref{eq22}) yields:
\be
p\tro=\frac{k_B\,T}{E[\rho]}\int d\sg\ros
+\frac{1}{2}\int d\sg\int d\sg'\,V(\sg,\sg')\ros\rho(\sg').
\label{eq35}
\ee
If $D(\sg,\sg')$ denotes the contact distance between two particles
of species $\sg$ and $\sg'$, we have $V_A(r;\sg,\sg')=0$ when
$r<D(\sg,\sg')$ so that~(\ref{eq33}) can always be rewritten~\cite{lovet,xu}:
\be
V(\sg,\sg')=-\epsilon(\sg,\sg')\frac{4\pi}{3}(D(\sg,\sg'))^3
\label{eq36}
\ee
where $\epsilon(\sg,\sg')>0$ characterizes the amplitude of the attractions.
For simplicity we will use the Lorentz-Berthelot mixing rules~\cite{deho,lero}:
\ba
D(\sg,\sg')=\{D(\sg,\sg)+D(\sg',\sg')\}/2\nonumber\\
\epsilon(\sg,\sg')=\{\epsilon
(\sg,\sg)\epsilon(\sg',\sg')\}^{1/2}
\label{eq37}
\ea
together with the additivity assumption, $D(\sg,\sg)=2\,R(\sg)$, $R(\sg)$
being the radius of a particle of species $\sg$, so that $v(\sg)=(4\pi/3)R^3(\sg)$.
Choosing $\sg=1$ as a reference species we may use $R(1)$ as length scale and
$\epsilon(1,1)$ as energy scale. Our basic (dimensionless) polydispersity
variable will hence be, $R(\sg)/R(1)$, which will be denoted here,
$\sg=R(\sg)/R(1)$, so that $\sg$ denotes here both a species and its associated
polydispersity variable, $R(\sg)/R(1)$. Below it will be convenient to consider various
particular cases of ~(\ref{eq36}-\ref{eq37}) by introducing two indices, $0\leq l\leq 1$ and $0\leq n\leq 1$,
and rewrite $D(\sg,\sg)=\sg^n D(1,1)$, $v(\sg)=\sg^{3n}v(1)$, $\epsilon(\sg,\sg)
=\sg^{2l}\epsilon(1,1)$ so that \{$l=0$, $n=0$\} corresponds to the
absence of polydispersity, \{$l=1$, $n=0$\} to polydispersity of the amplitude
or interaction strength only, \{$l=0$, $n=1$\} polydispersity in size only,
while \{$l=1$, $n=1$\} corresponds to both amplitude and size polydispersity.
Note that the amplitude polydispersity~\cite{stap} ($l\neq0$) includes the polydispersity
which originates from the species dependence of the range of the interaction
potential ($V_A(r;\sg,\sg')$). Finally, the present vdW-model is only a particular case of the general class of models considered by Gualtieri et al~\cite{gual}. It has, however, the advantage that it can be easily related to a realistic interaction potential via Eqs.~(\ref{eq33}) and~(\ref{eq36}). 

\section{The parent phase distribution}
\label{sec4}
To complete the description of our system we shall now specify, $\rho_0(\sg)
=\rho_0h_0(\sg)$, the distribution of the parent phases to be considered here.
The polydispersity distribution $h_0(\sg)$ is constrained by the fact that
it must be non-negative and normalized. We will assume moreover that the
values of $\sg=R(\sg)/R(1)$ are distributed continuously within $0<\sg<\infty$.
Of course, very large values of $\sg$ are unphysical but these will be
given a very small weight by requiring that $h_0(\sg)$ decays with $\sg$ in a
manner which is sufficiently rapid for all the moments of $h_0(\sg)$, $m_k^{(0)}
=\int_0^\infty d\sg\sg^k\,h_0(\sg)$, to exist. Moreover, we will restrict ourselves
to monomodal distributions~\cite{hunt} so that the systems considered here are polydisperse
generalizations of single component systems. Two often used candidates~\cite{elia,hunt}
are the Schulz-Zimm~(SZ) distribution, $h_0^{(SZ)}(\sg)=c\,\sg^{a}\exp(-b\sg)$,
and the log-normal~(LN) distribution, $h_0^{(LN)}(\sg)=c\,\exp\{-a\left(\ln(\sg/b)
\right)^2\}$. In each case the parameters \{$a$, $b$, $c$ \} are
determined by the normalization ($m_0^{(0)}=1$), the mean value ($m_1^{(0)}$)
and the polydispersity index, $I=m_2^{(0)}/(m_1^{(0)})^2$, of the distribution.
In what follows we will take $m_1^{(0)}=1$ so that the size of the reference
species, $R(1)$, is equal to the average value of $R(\sg)$ in the parent
phase distribution. Under these circumstances the SZ distribution can be written:
\be
h_0^{(SZ)}(\sg)=\frac{\alpha^\alpha}{\Gamma(\alpha)}\cdot\sg^{\alpha-1}\cdot\exp(-\alpha\sg)
\label{eq41}
\ee
where $\Gamma(\alpha)$ is the Euler gamma function of argument $0<\alpha<\infty$,
a parameter which determines the inverse width of the SZ distribution, or
equivalently, its polydispersity index, $I^{(SZ)}=1\,+1/\alpha$. The moments
of~(\ref{eq41}) are given by:
\be
m_k^{(SZ)}=\frac{1}{\alpha^k}\;\frac{\Gamma(\alpha+k)}{\Gamma(\alpha)}
\label{eq42}
\ee
which, when $k$ is an integer, can be rewritten (for $k\geq 2$) as:
\be
m_k^{(SZ)}=\Pi_{n=1}^{k-1}(1+\frac{n}{\alpha}).
\label{eq43}
\ee
For the LN distribution we have similarly:
\be
h_0^{(LN)}(\sg)=\frac{I}{\left(2\pi\ln I \right)^{1/2}}\cdot\exp
\left\{-\frac{\ln^2[I^{3/2}\,\sg]}{2\ln I}\right\}
\label{eq44}
\ee
where $I^{(LN)}=I$ while the moments of~(\ref{eq44}) are given by:
\be
m_k^{(LN)}=I^{k(k-1)/2}.
\label{eq45}
\ee
In both cases we have hence $m_0^{(0)}=m_1^{(0)}=1$ and $m_2^{(0)}=I=1+1/\alpha$. Note that
in the monodisperse limit, ($I\rightarrow1$ or $\alpha\rightarrow\infty$) both
these distributions reduce to the Dirac distribution, $\delta(\sg-1)$, centered
on the reference species ($\sg=1$). On the contrary, when $I$ becomes very large ($I>>1$) the above distributions become very wide, increasing hereby the importance of the larger particles (see Fig.1). As stated above, the presence of very large particles is unphysical and a realistic situation should be described by distributions which strictly vanish for $\sigma$ larger than some maximum value. To avoid such problems, below we will consider only values of $I$ in the range $1<I<2$. For such values the results obtained from the full distributions will be similar to those obtained from the physically truncated distributions (with a slightly different $I$-value).

\section{The solution method}
\label{sec5}
In what follows it will be convenient to use dimensionless variables.
Using $\epsilon(1,1)$ as energy scale and $R(1)$ as length scale,
$\sg=1$ being the reference species, we can introduce $t=k_B\,T/\epsilon(1,1)$,
$\overline\mu=\mu/\epsilon(1,1)$ as reduced temperature and chemical potential
while, if $v(1)=\frac{4\pi}{3}R^3(1)$ represents the volume of the reference
particle, for the reduced free-energy, pressure and average density we will
use, respectively, $\overline f=f\,v(1)/\epsilon(1,1)$, $\overline{p}=p\,v(1)/\epsilon(1,1)$,
$\eta=\rho\,v(1)$. The generalized moments ($m_k(\sg)=\sg^k$ but $k$ need not
be an integer) of the polydispersity distribution $h(\sg)$ will be written:
\be
m_k[h]=\int_0^\infty d\sg\,m_k(\sg)\,h(\sg)
\label{eq51}
\ee
or, shortly, $m_k=m_k[h]$ for a general distribution $h(\sg)$ and
$m_k^{(j)}=m_k[h_j]$ when $h_j(\sg)$ represents the distribution
of phase $j$ ($j=0,1,2$). Finally, as explained at the end of section~\ref{sec3},
$V(\sg,\sg')$ will be written:
\be
V(\sg,\sg')=V(1,1)(\sg\,\sg')^l\,\left(\frac{\sg^n+\sg'^n}{2}\right)^3
\label{eq52}
\ee
with $V(1,1)=-8\ \epsilon(1,1)v(1)$, 
where the value of $l$ controls the amplitude polydispersity while the value
of $n$ controls the size polydispersity. In terms of these variables we can rewrite
the excess part of eqs.~(\ref{eq34}), needed for~(\ref{eq214}), as:
\ba
\overline\mu_{ex}(\sg,t,\eta,[h])=-t\ \ln(1-\eta\,m_{3n})
+\frac{t\eta\,m_0 m_{3n}(\sg)}{1-\eta\,m_{3n}}
\nonumber\\
-\eta\left\{m_{l+3n}(\sg)m_l+3m_{l+2n}(\sg)m_{l+n} 
+3m_{l+n}(\sg)m_{l+2n} +m_{l}(\sg)m_{l+3n}\right\} \nonumber\\
\label{eq53}
\ea
while~(\ref{eq35}) becomes:
\be
\overline p(t,\eta,[h])=\frac{t\eta\,m_0}{1-\eta\,m_{3n}}
-\eta^2(m_{l}m_{l+3n}+3\,m_{l+n}m_{l+2n})
\label{eq54}
\ee
where it is seen that because of the polynomial character of~(\ref{eq52}) these
expressions involve only a finite number of moments $m_k$, \{$k=0,3n,l,l+n,
l+2n,l+3n$\}. Note that the exact number of moments depends on the values
of $l$ and $n$ (for ex. when $l=0$, $k=\{0,l\}$ or $k=\{3n,l+3n\}$ represent one
and the same moment, resp. $k=0$ and $k=3n$). In the present context we can
thus rewrite~(\ref{eq216}) as:
\be
h_1(\sg)=h_0(\sg)\,H(t,\eta_0,\eta_1,\eta_2;\{m_k(\sg)\},
\{m_k^{(1)}\},\{m_k^{(0)}\})
\label{eq55}
\ee
where the $\{m_k^{(2)}\}$ have been eliminated in favor of $\{m_k^{(0)}\}$
and $\{m_k^{(1)}\}$ by using~(\ref{eq212}):
\be
m_k^{(2)}=\frac{v_2-v_1}{v_0-v_1}m_k^{(0)}+ \frac{v_2-v_0}{v_1-v_0}m_k^{(1) }.
\label{eq56}
\ee
Although eq.~(\ref{eq55}) can be solved directly, as was done in~\cite{xu},
here we will take
 advantage of the structure of~(\ref{eq55}) to transform this integral
equation for $h_1(\sg)$ into a set of moment relations~\cite{gual}:
\be
m_{k'}^{(1)}=\int d\sg m_{k'}(\sg)\,h_0(\sg)\,H(t,\eta_0,\eta_1,\eta_2;\{m_k(\sg)\},
\{m_k^{(1)}\},\{m_k^{(0)}\})
\label{eq57}
\ee
where both $k'$ and $k$ run through the finite set of moments $\{0,3n,l,l+n,l+2n,l+3n\}$.
Note that for $k'=0$, eq.~(\ref{eq57}) corresponds to~(\ref{eq217}) while~(\ref{eq218})
becomes here:
\ba
\frac{t\eta_1\,m_0^{(1)}}{1-\eta_1\,m_{3n}^{(1)}}
-\eta_1^2(m_{l}^{(1)}m_{l+3n}^{(1)}+3\,m_{l+n}^{(1)}m_{l+2n}^{(1)})=
\frac{t\eta_2\,m_0^{(2)}}{1-\eta_2\,m_{3n}^{(2)}}
\nonumber\\
-\eta_2^2(m_{l}^{(2)}m_{l+3n}^{(2)}+3\,m_{l+n}^{(2)}m_{l+2n}^{(2)}) \nonumber\\
\label{eq58}
\ea
together with the normalization condition $m_{0}^{(0)}=m_{0}^{(1)}=m_{0}^{(2)}=1$. For any given $t$, $\eta_0$
and $h_0(\sg)$ (from which the $m_k^{(0)}$ can be determined)
the seven unknowns \{$\eta_1$,$\eta_2$,$m_{3n}^{(1)}$,$m_{l}^{(1)}$,
$m_{l+n}^{(1)}$,$m_{l+2n}^{(1)}$,$m_{l+3n}^{(1)}$\} can then be obtained
by solving the system of seven equations~(\ref{eq57}) and~(\ref{eq58}) for
\{$k'=0,3n,l,l+n,l+2n,l+3n$\}. When this result is substituted in~(\ref{eq55})
we obtain $h_1(\sg)$ and from~(\ref{eq212}) we obtain then $h_2(\sg)$. This
then completely solves the two-phase coexistence problem for the
present vdW-model. In the monodisperse limit we have $m_k^{(j)}=1$, for
all $k$ and $j$ values, and~(\ref{eq57}-\ref{eq58}) reduce then to the
usual vdW equations for the binodal of the reference species. To solve~(\ref{eq57}-\ref{eq58})
we have used an iterative process by starting from an initial guess of the solution
(usually the monodisperse
case). When using moreover a globally convergent Newton-Raphson method~\cite{nume}
a (convergent) solution of~(\ref{eq57}-\ref{eq58}) can easily be obtained
even close to the critical points. The latter can hence be determined
directly from the binodals but they can also be obtained by alternative
procedures. One such procedure which is based on the stability analysis
of section~\ref{sec2p3} will be detailed in the next section whereas
here we will focus on an alternative procedure based on the determination
of the so-called cloud-point (C) and shadow (S) curves. These curves provide
envelopes for the binodals. They can be obtained from~(\ref{eq57}-\ref{eq58})
by considering a situation of insipient phase separation whereby phase 1 is
present only in infinitesimal amounts. Returning to section~\ref{sec2p2}
this situation is seen to correspond to $x_1\rightarrow 0$ or
$v_2\rightarrow v_0$ with $v_1$ and $v_2$ finite. From~(\ref{eq212})
and~(\ref{eq56}) it is seen that this implies $h_2(\sg)\rightarrow h_0(\sg)$
and $m_k^{(2)}\rightarrow m_k^{(0)}$. These curves are hence solutions of
(cf.~(\ref{eq55},\ref{eq57}-\ref{eq58})):
\be
h_1(\sg)=h_0(\sg)\,A_{12}(t,\eta_0=\eta_2,\eta_1;\{m_k(\sg)\},
\{m_k^{(1)}\},\{m_k^{(0)}\})
\label{eq59}
\ee    
\be
m_{k'}^{(1)}=\int_{0}^{\infty} d\sg m_{k'}(\sg)\,h_0(\sg)\,A_{12}(t,\eta_0=\eta_2,\eta_1;
\{m_k(\sg)\},
\{m_k^{(1)}\},\{m_k^{(0)}\})
\label{eq510}
\ee
\ba
\frac{t\eta_1\,m_0^{(1)}}{1-\eta_1\,m_{3n}^{(1)}}
-\eta_1^2(m_{l}^{(1)}m_{l+3n}^{(1)}+3\,m_{l+n}^{(1)}m_{l+2n}^{(1)})=
\frac{t\eta_0\,m_0^{(0)}}{1-\eta_0\,m_{3n}^{(0)}}
\nonumber\\
-\eta_0^2(m_{l}^{(0)}m_{l+3n}^{(0)}+3\,m_{l+n}^{(0)}m_{l+2n}^{(0)}) \nonumber\\
\label{eq511}
\ea
where we took moreover into account in~(\ref{eq59}-\ref{eq510}) that,
$H\rightarrow A_{12}$, when $v_2\rightarrow v_0$ (cf.~(\ref{eq215}-\ref{eq216}).
The solution to~(\ref{eq59}-\ref{eq511}) corresponding to the majority phase
$2$ yields then the so-called C-curve while the solution for the minority
phase $1$ yields the S-curve.

\section{Critical point analysis of the vdW free-energy}
\label{sec6}
As explained in section~\ref{sec2p3}, a free-energy density $f(T,[\rho])$
will be globally stable if it remains a convex functional of $\ros$ for
any change, $\ros\rightarrow\ros+\delro(\sg)$. It then must satisfy
eq.~(\ref{eq219}). The same free-energy will be locally stable when~(\ref{eq220})
holds, i.e. when the first functional variation $\delta^k f(T,[\rho])$
of $f(T,[\rho])$ which is non-zero corresponds to an even value of
$k\geq2$. We then have \{$\delta^2 f>0$\} for a (ordinary) stable state;
\{$\delta^2 f=0$, $\delta^3 f=0$, $\delta^4 f>0$\} for a (ordinary)
critical state, \{$\delta^2 f=\delta^3 f=\delta^4 f=\delta^5 f=0$,
$\delta^6 f>0$\} for a tricritical state, etc. Here we will limit ourselves
to the equations determining the ordinary critical states of the vdW
free-energy of section~\ref{sec3}. From~(\ref{eq31}) and~(\ref{eq222})
we obtain ($\beta=1/k_B\,T$; $\overline{v}(\sigma_{n})=v(\sigma_{n})/E[\rho]$):
\be
\beta K_2(\sg_1,\sg_2;T,[\rho])=\frac{\delta(\sg_1-\sg_2)}{\rho(\sg_1)}
+\beta V(\sg_1,\sg_2)+[\overline{v}(\sg_1)+\overline{v}(\sg_2)]
+\rho\ \overline{v}(\sg_1)\ \overline{v}(\sg_2)
\label{eq61}
\ee
\ba
\beta K_3(\sg_1,\sg_2,\sg_3;T,[\rho])=-\frac{\delta(\sg_1-\sg_2)
\delta(\sg_2-\sg_3)}{\rho(\sg_1)\rho(\sg_2)}
+2\rho\ \overline{v}(\sg_1)\ \overline{v}(\sg_2)\ \overline{v}(\sg_3)
\nonumber\\
+[\overline{v}(\sg_1)\ \overline{v}(\sg_2)+\overline{v}(\sg_2)\ \overline{v}(\sg_3)+\overline{v}(\sg_3)\ \overline{v}(\sg_1)]
\label{eq62}
\ea
\ba
\beta K_4(\sg_1,\sg_2,\sg_3,\sg_4;T,[\rho])=2
\frac{\delta(\sg_1-\sg_2)}{\rho(\sg_1)}
\frac{\delta(\sg_2-\sg_3)}{\rho(\sg_2)}
\frac{\delta(\sg_3-\sg_4)}{\rho(\sg_3)}
\nonumber\\
+6\rho\ \overline{v}(\sg_1)\ \overline{v}(\sg_2)\ \overline{v}(\sg_3)\ \overline{v}(\sg_4)
+2\ [\overline{v}(\sg_1)\ \overline{v}(\sg_2)\ \overline{v}(\sg_3)
+\overline{v}(\sg_1)\ \overline{v}(\sg_2)\ \overline{v}(\sg_4)
\nonumber\\
+\overline{v}(\sg_1)\ \overline{v}(\sg_3)\ \overline{v}(\sg_4)
+\overline{v}(\sg_2)\ \overline{v}(\sg_3)\ \overline{v}(\sg_4)]
\label{eq63}
\ea
where it is seen that $K_4$ is positive definite while $K_3$ and $K_2$ can
vanish because the volume $v(\sg)$ of species $\sg$ and $E[\rho]$ of~(\ref{eq32})
are positive while $V(\sg_1,\sg_2)$ of~(\ref{eq36}) is negative. We now turn to eq.~(\ref{eq224})
which can be cast into an eigenvalue form~\cite{kinc,cues} by rewriting
$\delro(\sg)$ as $\delro(\sg)=\ros\cdot e(\sg)$ where $\ros=\rho h(\sg)$
and $e(\sg)$ is as yet unknown. Using moreover the dimensionless variables
introduced in section~\ref{sec5} we can recast~(\ref{eq224}) and~(\ref{eq61})
in the form:
\be
\int d\sg_1 d\sg_2\,e(\sg_{1})\{\left(h(\sg_1)h(\sg_2)\right)^{1/2}\delta(\sg_1-\sg_2)
+h(\sg_1)\,C_2(\sg_1,\sg_2;t,\eta)\,h(\sg_2)\}e(\sg_2)=0
\label{eq64}
\ee
where:
\ba
C_2(\sg_1,\sg_2;t,\eta)&=A\left(m_{3n}(\sg_1)m_0(\sg_2)
+m_0(\sg_1)m_{3n}(\sg_2)  \right)+A^2\,m_{3n}(\sg_1)\,m_{3n}(\sg_2)\nonumber\\
&-B\{m_{l+3n}(\sg_1)m_{l}(\sg_2)+3m_{l+2n}(\sg_1)m_{l+n}(\sg_2)
\;\;\;\;\;\nonumber\\
&\;\;\;\;\;+3\,m_{l+n}(\sg_1)m_{l+2n}(\sg_2)+m_{l}(\sg_1)m_{l+3n}(\sg_2)\}
\label{eq65}
\ea    
together with the shorthand notations, $A=\eta/(1-\eta m_{3n})$, and $B=\eta/t$.
For convenience we rewrite~(\ref{eq65}) briefly as:
\be
C_2(\sg_1,\sg_2;t,\eta)=\sum_{k,k'}c_{k k'}(t,\eta) m_k(\sg_1)m_{k'}(\sg_2)
\label{eq66}
\ee
where the $c_{kk'}$ can be deduced by identification of~(\ref{eq66})
with~(\ref{eq65}). From~(\ref{eq66}) it is obvious that $C_2(\sg_1,\sg_2;t,\eta)$
is completely embedded in the subspace spanned by the vdW-moments
\{$m_k(\sg)$; $k=0,3n,l,l+n,l+2n,l+3n$\}. To solve~(\ref{eq64}) it will hence
be sufficient to restrict $e(\sg)$ to the same subspace~\cite{cues}, i.e. $e(\sigma)=\sum_{k}e_{k}m_{k}(\sigma)$ . A solution of the
homogeneous equation~(\ref{eq64}) will hence exist provided the determinant
of the matrix, 
$\underline{\underline{M}}+\underline{\underline{M}}
\cdot\underline{\underline{C}}\cdot\underline{\underline{M}}$ vanishes. We can rewrite this matrix as 
$\underline{\underline{M}}\cdot\underline{\underline{A}}$, 
with $\underline{\underline{A}}=\underline{\underline{I}}
+\underline{\underline{B}}$ and
$\underline{\underline{B}}=\underline{\underline{C}}
\cdot\underline{\underline{M}}$. Here, $I_{kk'}=\delta_{kk'}^{Kr}$ is the
unit matrix, $C_{kk'}=c_{kk'}(t,\eta)$ the matrix of coefficents of~(\ref{eq66}),
and $M_{kk'}=m_{k+k'}$ a matrix involving the moments of $h(\sg)$. (Note
that $m_k(\sg)m_{k'}(\sg)=m_{k+k'}(\sg)$). Since $det|M_{kk'}|\neq0$,  the equation
of the spinodal (SP) can be written in various forms, e.g. :
\be
det|I_{kk'}+B_{kk'}|=0,\;\;\;B_{kk'}=\sum_{k''}c_{kk''}m_{k''+k'}
\label{eq67}
\ee
whereas the corresponding solution of~(\ref{eq64}) reads (up to a proportionality
factor):
\be
e(\sg)=\sum_k a_{k'k}m_k(\sg)
\label{eq67a}
\ee
where $a_{k'k}$ is the co-factor (algebraic minor) of $A_{k'k}=\delta_{k'k}^{Kr}
+B_{k'k}$ (note that the r.h.s. of~(\ref{eq67a}) is in fact independent of the value
of $k'$). Having found the spinodal (SP)
we now turn to the stability condition (ST), eq.~(\ref{eq225}). Using~(\ref{eq62})
and writing $\delro_0(\sg)=\ros\,e(\sg)$, eq.~(\ref{eq225}) can be rewritten
($E=E[\rho]$):
\ba
\int d\sg\,h(\sg)\,e^3(\sg)=3\ (\frac{\rho}{E})^2
\{ \int d\sg\,h(\sg)\,e(\sg)\}
\{ \int d\sg\,h(\sg)\,e(\sg)\,v(\sg)\}^2\nonumber\\
+2(\frac{\rho}{E})^3
\{ \int d\sg\,h(\sg)\,e(\sg)\,v(\sg)\}^3
\label{eq68}
\ea
which on choosing $k'=0$ in~(\ref{eq67a}) and substituting the result
into~(\ref{eq68}) yields:
\ba
\sum_{k,k',k''} a_{0k}a_{0k'}a_{0k''}\,m_{k+k'+k''}=
&3A^2(\sum_k a_{0k}m_k)(\sum_{k'} a_{0k'}m_{k'+3n})^2\nonumber\\
&+2A^3(\sum_k a_{0k}m_{k+3n})^3\;\;\;\;\;
\label{eq69}
\ea
where, as before, $A=\eta/(1-\eta m_{3n})$. Finally, in the vdW model
eq.~(\ref{eq226}) is always satisfied because $K_4$ of~(\ref{eq63}) is positive
definite.

In summary, those thermodynamic states \{$t,\eta,h(\sg)$\} for which
$\delta^2 f>0$ are stable while the marginally stable states
($\delta^2 f=0$) obey the spinodal condition~(\ref{eq67}). The critical states
are stable marginal states, i.e. they must obey both eq.~(\ref{eq67})
and~(\ref{eq69}). Note that these equations are algebraic equations involving
t, $\eta$ and moments of the type $m_k$, $m_{k+k'}$ and $m_{k+k'+k''}$ where \{$k,k',k''$\}
cover the finite set of values \{$0,3n,l,l+n,l+2n,l+3n$\} appearing in
the excess free-energy. 

\section{Some typical phase diagrams}
\label{sec7}
Having obtained the basic equations which govern the phase behavior of the present
polydisperse vdW-fluid we now consider a few explicit examples. In order to
completely characterize a thermodynamic situation we must specify ($t,\eta_0$),
the parent phase distribution $h_0(\sg)$, and the values of ($l,n$) to be used in~(\ref{eq52}).
In our study we have found no qualitative differences between the two types (SZ
or LN) of parent phase distributions considered in section~\ref{sec4}. All features
found for one type can be found for the other type by changing the value of the
polydispersity index $I$. On the contrary, the influence of the ($l,n$) values is much
more pronounced, as we now explain.

\subsection{Interactions with size polydispersity only}
\label{sec7p1}
If we remove the amplitude polydispersity ($l=0$) and keep only the size polydispersity
($n=1$) of the interaction potential the present vdW model is governed by the following
four moments \{$m_0,m_1,m_2,m_3$\} , i.e. $k=0,1,2,3$ (see section~\ref{sec5}).
The equations~(\ref{eq57}-\ref{eq58}) governing the binodals constitute then a system of five equations which can be solved numerically and similarly for the C- and S-curves (see ~(\ref{eq510}-\ref{eq511})), as explained in section~\ref{sec5}. Returning
to~(\ref{eq54}) it is seen that in the present case the effect of the polydispersity
($I\neq 1$) on the repulsions (or excluded volume) is stronger ($\eta\rightarrow\eta m_3$)
than its effect on the attractions ($4\eta^2\rightarrow\eta^2(m_3+3m_1 m_2$))
because the moments $m_k$ are increasing functions of $I=1+1/\alpha>1$
(cf.~(\ref{eq43}) and~(\ref{eq45})). This situation is hence unfavorable to phase separation
when compared with the (strictly) monodisperse case ($I=1$). This can be seen in the
phase diagram shown in Fig.2. The result shown corresponds to a SZ parent phase of
relatively modest polydispersity index $I=1.04$ ($\alpha=25$). Note that at present~\cite{ref7}
it is current practice to consider colloidal dispersions with $I<1.05$ as (approximately)
monodisperse. It is seen in the figure that, when compared to the (strictly) monodisperse
case ($I=1$), the coexistence  region is shifted to lower temperatures. As indicated,
there exists now a different binodal for each value of $\eta_0$, (instead of the unique
binodal when $I=1$). These binodals are truncated on the high-temperature side and
fill the space between the cloud-point(C) and shadow (S) curves. The high-density phase
is shifted towards lower densities whereas the low-density phase is only slightly
shifted towards higher densities. Note also that the polydispersity distributions,
$h_1(\sg)$ and $h_2(\sg)$, change with $T$ along the binodals (see Fig.3). The
S-curve is situated in the interior of the C-curve and is tangent to the latter at the
commun maximum of both curves. This commun maximum corresponds to a critical
point where the two phases become identical. The binodal through the critical
point is the only untruncated binodal. The critical point itself is shifted to lower
temperatures and lower densities when compared to the monodisperse case. When
increasing the value of $I$ all these shifts increase monotonically without
modifying the overall aspect of the phase diagram. As far as we are aware, phase diagrams of this type have not yet been found experimentally but should be observable in suitably prepared colloidal dispersions.

\subsection{ Interactions with amplitude polydispersity only}
\label{sec7p2}
If we remove the size polydispersity ($n=0$) but keep the amplitude polydispersity
($l\neq0$, say $l=1$) the present vdW model is governed by two moments \{$m_0,m_1$\},
i.e. $k=0,1$ and eqs.~(\ref{eq57}-\ref{eq58}) reduce to a system of three equations.
In this case the polydispersity does only affect the attractions (see, $4\eta^2
\rightarrow\eta^2\cdot 4m_1^2$, in eq.~(\ref{eq54})), a situation which is favorable
to phase separation. This is seen in Fig.4 where the phase coexistence region is shown
to be shifted to higher temperatures. Note also that the S-curve has moved now
to higher densities and also moved partly outside of the interior of the C-curve. As a result the
intersection of the C- and S-curves, i.e. the critical point, is encountered now
below the temperature $(t_{m})$ corresponding to the commun maximum of the C-
and S-curves. As seen in Fig.4b this allows for a re-entrant behavior of the low
density phase. The same behavior is found for any $l>0$, whatever small. Phase diagrams of this type have been found for some polymeric systems~\cite{clar}.

\subsection{Size and amplitude polydispersity}
\label{sec7p3}
In a realistic situation we expect $n=1$ and $l\neq 0$. This increases the number of
moments (up to a maximum of six, five for $l=1$) but it is easily verified that within
the present vdW model the attractions are always more strongly affected by the
polydispersity than the repulsions and will hence always favor the phase separation.
Examples of the resulting phase diagrams are shown in Fig.5. Note that the modifications introduced into the phase diagram by the amplitude polydispersity (compare Figs.2 and 5) could be used to probe the presence or absence of an amplitude polydispersity in the interaction between the colloidal particles being studied.

\section{Polydispersity induced critical points}
\label{sec8}
In the above we have limited ourselves to the case where only two phases coexist.
We have seen that the phase diagram is considerably modified by the polydispersity,
in particular in the region of the critical point. When the temperature is lowered,
starting from the critical temperature, one expects to encounter a region where
three, four, etc. phases coexist. Because only fluid phases are involved we expect
the transition from the two-phase to the three-phase region to proceed through a
second critical point. Since such a second critical point is absent from the ordinary
monodisperse vdW fluid it can be termed polydispersity induced. In the present section
we will describe how these polydispersity induced critical points can be found in
a systematic way. Critical points can be obtained by following different routes,
two of which have already been illustrated in section~\ref{sec7}. It was seen there
that critical points can be found by either looking for untruncated binodals or for
intersections of the C- and S- curves. When the critical region is approximately known
these routes are easily executed but this method becomes unpractical when no a priori
knowledge about their location is available, as is the case here for all the polydispersity
induced critical points. The method to be followed here will therefore be based on the
stability criteria of section~\ref{sec2p3}, as adapted to the vdW case in 
section~\ref{sec6}. When following this route the critical points can be found by
looking for intersections of the spinodal (SP) and the stability (ST) curves. In the present
context this amounts to look for thermodynamic states ($t,\eta$) which, for a given
$h(\sg)=h_0(\sg)$, satisfy both eq.~(\ref{eq67}) and~(\ref{eq69}). In Fig.6 we show how
the various routes co\"{\i}ncide for the ordinary, high-temperature, vdW critical
point. Some of the remaining polydispersity induced critical points obtained from the present route
are shown in Fig.7.

 It is seen there that when increasing the polydispersity index $I$, the number of
polydispersity induced critical points gradually increases. It is also seen that
at the same time the region of spinodal instability ($\delta^2 f<0$ for $\delta\rho(\sigma)=\delta\rho_{0}(\sigma)$) rapidly invades
the high density region of the phase diagram. As a consequence, some of the
polydispersity induced transitions and critical points could still be metastable
with respect to the solid phases expected to become stable in this high density
region.

\section{Conclusions}
\label{sec9}
In the present study we have presented phase diagrams for fluids composed of spherical
particles with a monomodal size distribution. These phase diagrams have been obtained
on the basis of the van der Waals approximation for the free-energy of a polydisperse
fluid. Interaction potentials with both size and amplitude polydispersity have been
considered. It has been found that the largest modifications to the phase diagram of
the polydisperse fluid, as compared to its monodisperse counterpart, result from the
amplitude polydispersity of the interaction potential. These modifications should already
be observable in colloidal dispersions with a relatively small polydispersity index $I$
(e.g. $1.01\leq I\leq 1.05$). For larger, but still modest values of $I$ (e.g. $1.1 \leq I \leq 1.5$) a second, polydispersity induced, critical point is found signaling the onset of a three phase coexistence. When the value of $I$ is increased still further, high-order coexistences are found but these are located in the high-density region of the phase diagram where the fluid phases considered here are expected to become metastable with respect to the solid phases.
\vspace{0.3truecm}

Acknowledgements
\newline
M.B. acknowledges financial support from the F.N.R.S.

\pagebreak
\renewcommand{\baselinestretch}{1.5}

\pagebreak
\noindent{\bf Figure Captions}\par
\vspace{1truecm}
\noindent
{\bf FIG. 1.} Two examples of the Schulz-Zimm (SZ) (solid lines) and log-normal (LN) (dotted lines) parent phase distributions, $h_0(\sigma)$, considered in this work (see ~(\ref{eq41}-\ref{eq44})). The values of the polydispersity index ($I$) are, $I=1.04$ and $I=4/3$ (as indicated). It is seen that for the values of $I$ considered here ($1<I<2$) these two distributions are rather similar. In particular, both are very small for the larger $\sigma$-values. Note also that the average value of $\sigma$ ($m_{1}^{(0)}=1$) does not coincide with the value of $\sigma$ for which $h_0(\sigma)$ reaches its maximum.
\par
\vspace{0.3truecm}
\noindent
{\bf FIG. 2.} An example of a phase diagram (in (a) the reduced temperature ($t=k_B\,T/\epsilon(1,1)$)-reduced density ($\eta=\rho\,v(1)$) plane and (b) the reduced pressure ($\overline{p}=pv(1)/\epsilon(1,1)$)-reduced temperature ($t$) plane) for the SZ parent phase distribution ($I=1.04$) shown in Fig.1 and the vdW-model ($l=0,n=1$) with only a size polydispersity.  For comparison the monodisperse phase diagram ($I=1$) (dashed line) has also been included. Shown are, the C-curve (full dots), the S-curve (open dots), and three binodals (full lines) for resp. $\eta_{0}=0.15$, $\eta_{0}=0.35$ and $\eta_{0}=\eta_{crit}=0.2852$ together with a tie line (dotted lines of (a)) corresponding to the upper temperature of coexistence (the lowest tie line shown belongs to the $\eta_0=0.15$ binodal). The critical point is indicated by a square (filled for $I=1.04$ and open for $I=1$). Note also that in the $p-t$ diagram (b) the C- and S-curves coincide (with $\eta_1<\eta_2$ on the upper branch and $\eta_1>\eta_2$ on the lower branch).
\par
\vspace{0.3truecm}
\noindent
{\bf FIG.3.} Evolution of the polydispersity distributions of the coexisting phases, $h_1(\sigma)$ and $h_2(\sigma)$, with the temperature along  the critical binodal ($\eta_0=\eta_{crit}=0.2852$) of Fig.2. Shown are : $h_1(\sigma)$ (full line, $t=0.9$; dash-dot line, $t=1.12$) and $h_2(\sigma)$ (dotted line, $t=0.9$; full dots, $t=1.12$).
\par
\vspace{0.3truecm}
\noindent
{\bf FIG.4.} The same as Fig.2 but for the vdW-model ($l=1$, $n=0$) with amplitude polydispersity only. The case shown corresponds to a SZ parent phase distribution with $I=1.02$. The three binodals shown correspond to $\eta_0=0.25$ (outer), $\eta_0=0.45$ (inner) and the critical binodal $\eta_0=\eta_{crit}=0.346$ (middle). Note that in the $p-t$ diagram the critical point corresponds to the maximum of the pressure but not to the maximum of the temperature ($t_m$) for coexistence. The latter can also be seen in the $t-\eta$ diagram. This implies a re-entrant behaviour for $t_{crit}<t<t_m$.
\par
\vspace{0.3truecm}
\noindent
{\bf FIG.5.} The same as Fig.2 but for the vdW-model ($l=1$, $n=1$) with both size and amplitude polydispersity. The case shown corresponds to a SZ parent phase distribution with $I=1.01$. The three binodals shown correspond to $\eta_0=0.25$ (outer), $\eta_0=0.4$ (inner) together with the critical binodal $\eta_0=\eta_{crit}=0.3338$ (middle). Note that this phase diagram is globally similar to the one of Fig.4 but shifted to higher temperatures although the value of $I$ is smaller here. The phase diagrams are very sensitive to the total amount of polydispersity present, i.e. to the values of $l$ and $n$.
\par
\vspace{0.3truecm}
\noindent
{\bf FIG.6.} The critical point shown in Fig.5 as determined by three different routes : 1) the maximum of the untruncated critical ($\eta_0=\eta_{crit}$) binodal(B), 2) the intersection of the C- and S-curves and 3) the intersection of the spinodal (SP) and stability (ST) curves. Note that the SP-curve is tangent to the C-curve at the critical point ($t_{crit}=1.2620$, $\eta_{crit}=0.3338$).
\par
\vspace{0.3truecm}
\noindent
{\bf FIG. 7.} Evolution of the number of critical points for a vdW fluid
with both size and amplitude polydispersity ($l=1$, $n=1$) and a LN parent phase distribution of increasing polydispersity
( (a): $I=1.1$, (b): $I=4/3$). Similar results can be  obtained for the
SZ
distribution. The critical points (full dots) correspond to the intersection of
the
SP (full line) and  the ST (dashed line) curves. Note that
when
$I$ increases the number of critical points increases (one for
$1<I<1.09$, two for $1.09<I<1.5$, etc.)
while
at the same time the spinodal region  invades the high density 
portion of the reduced  temperature ($t$)-average packing 
fraction ($\eta^*=\eta m_3$)
plane (here $m_3=I^{3}$
is the third moment of the LN distribution).   

\newpage
\thispagestyle{empty}
\hspace{5cm}
{\Large
\textbf{
Fig. 1}
, Bellier-Castella et al., JCP}
\\[2cm]
\Large
\begin{center}
\unitlength1mm
\begin{picture}(140,140)
\put(0,0){\makebox(140,140)[b]
{\epsfysize140mm \leavevmode \epsffile{fig1.epsi} }}
\end{picture}
\end{center}

\newpage
\thispagestyle{empty}
\hspace{5cm}
{\Large
\textbf{
Fig. 2a}
, Bellier-Castella et al., JCP}
\\[2cm]
\Large
\begin{center}
\unitlength1mm
\begin{picture}(140,140)
\put(0,0){\makebox(140,140)[b]
{\epsfysize140mm \leavevmode \epsffile{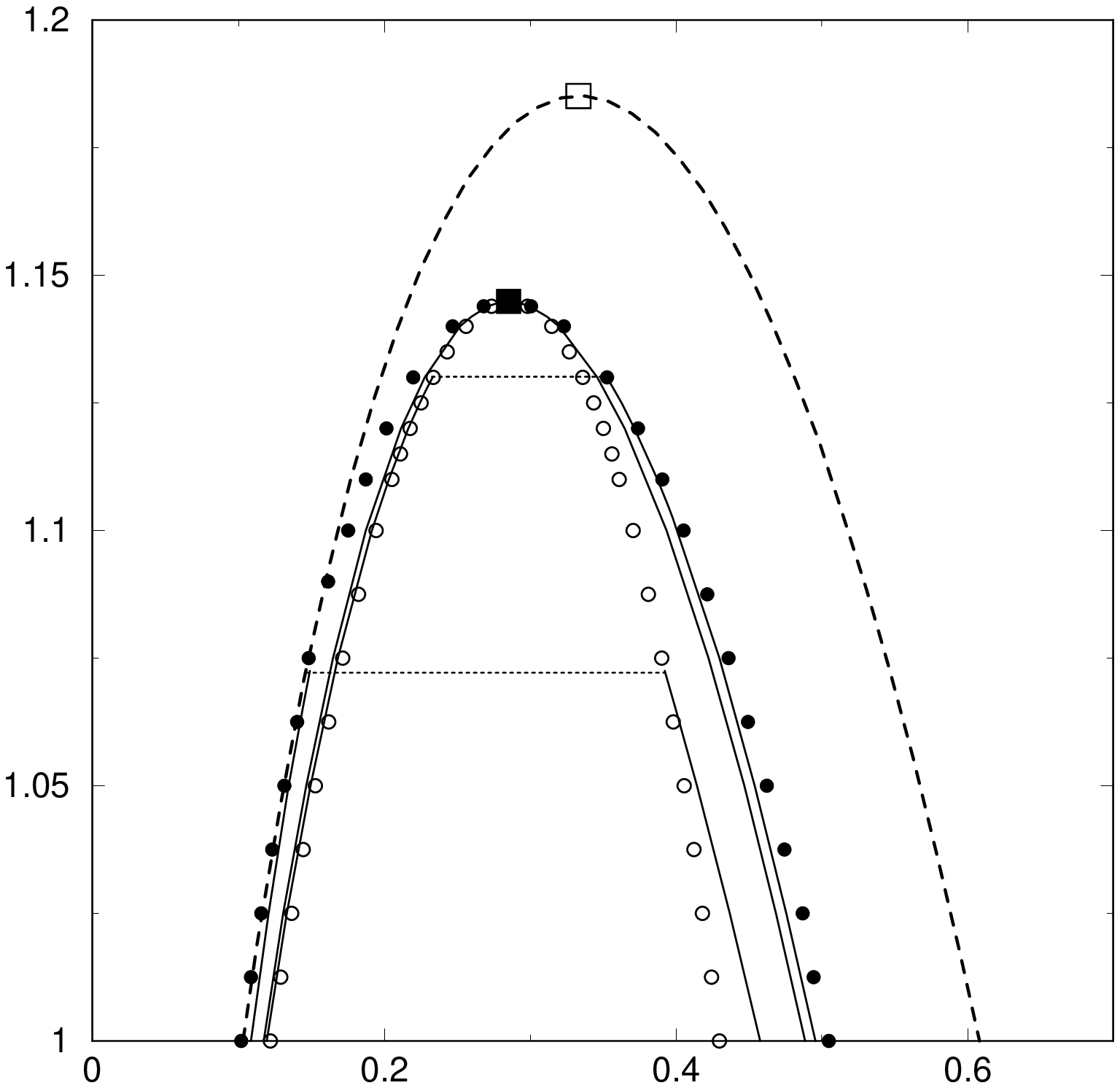} }}
\end{picture}
\end{center}

\vskip -10 cm
\hskip 0 cm
{\Large
$t$}

\vskip 8. cm
\hskip 7.5 cm
{\Large
$\eta$}

\newpage
\thispagestyle{empty}
\hspace{5cm}
{\Large
\textbf{
Fig. 2b}
, Bellier-Castella et al., JCP}
\\[2cm]
\large
\begin{center}
\unitlength1mm
\begin{picture}(140,140)
\put(0,0){\makebox(140,140)[b]
{\epsfysize140mm \leavevmode \epsffile{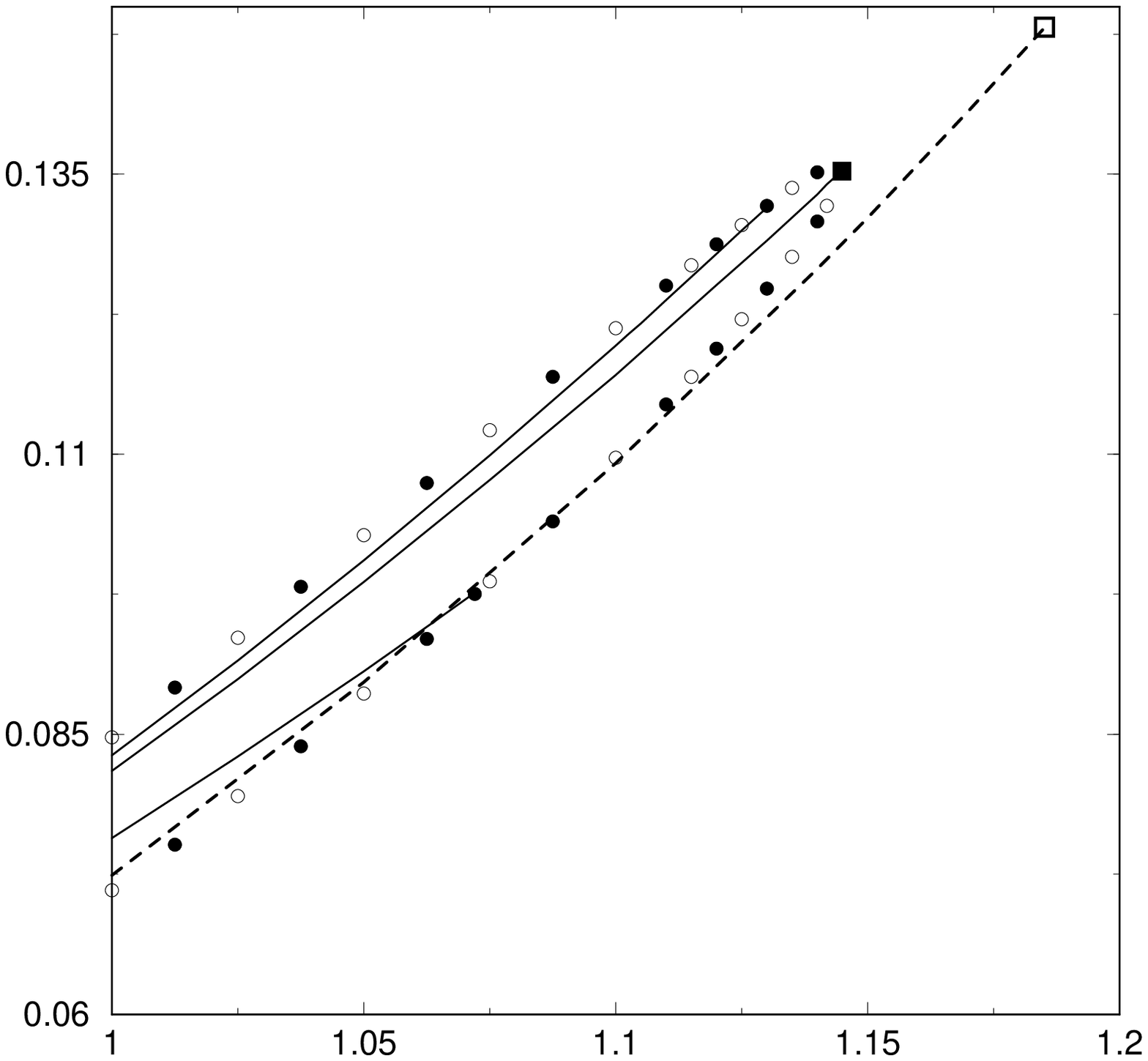} }}
\end{picture}
\end{center}

\vskip -11 cm
\hskip 0. cm
{\Large
$\overline{p}$}

\vskip 9.5 cm
\hskip 7.5 cm
{\Large
$t$}

\newpage
\thispagestyle{empty}
\hspace{5cm}
{\Large
\textbf{
Fig. 3}
, Bellier-Castella et al., JCP}
\\[2cm]
\large
\begin{center}
\unitlength1mm
\begin{picture}(140,140)
\put(0,0){\makebox(140,140)[b]
{\epsfysize140mm \leavevmode \epsffile{fig3c.epsi} }}
\end{picture}
\end{center}

\newpage
\thispagestyle{empty}
\hspace{5cm}
{\Large
\textbf{
Fig. 4a}
, Bellier-Castella et al., JCP}
\\[2cm]
\large
\begin{center}
\unitlength1mm
\begin{picture}(140,140)
\put(0,0){\makebox(140,140)[b]
{\epsfysize140mm \leavevmode \epsffile{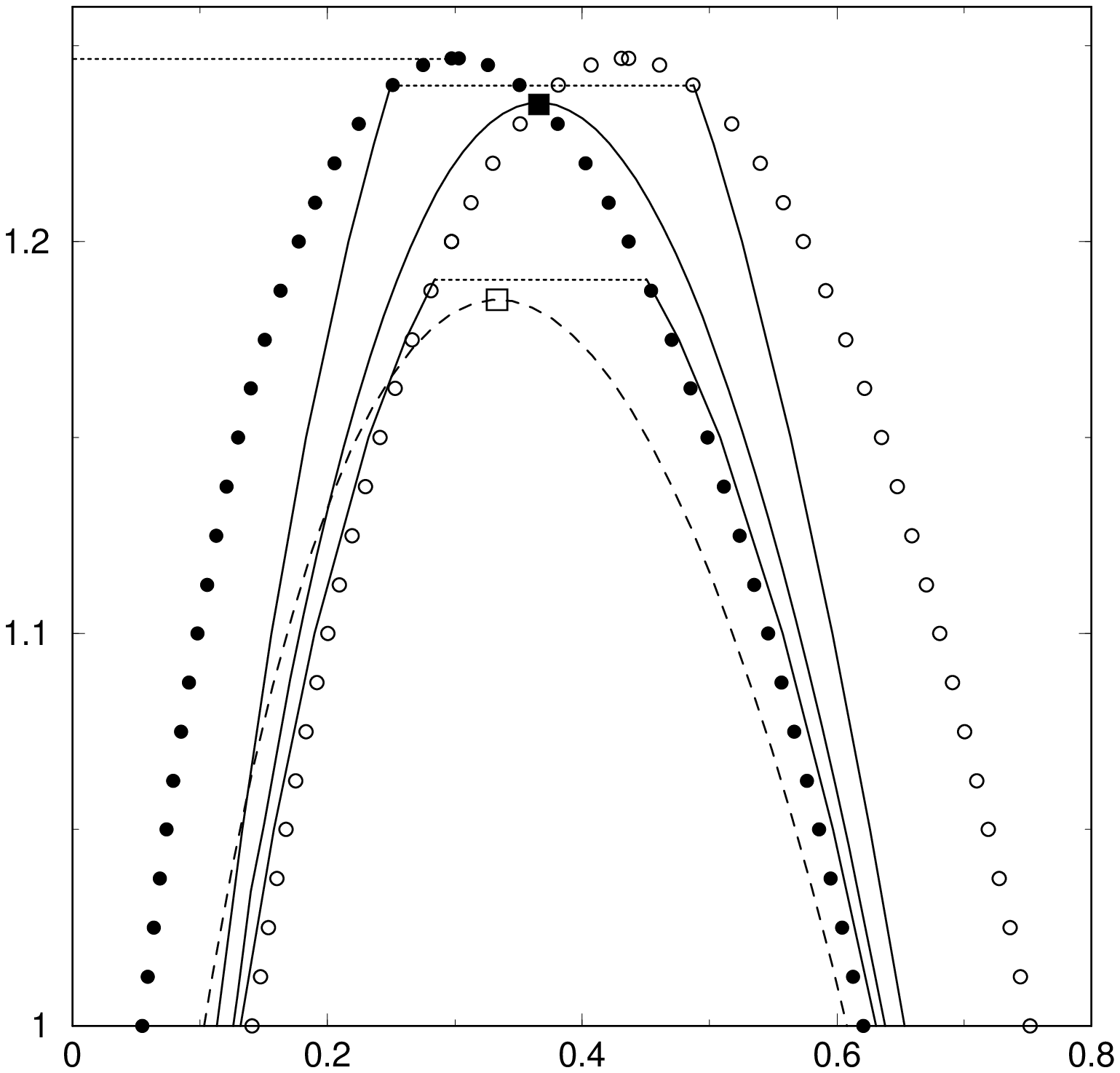} }}
\end{picture}
\end{center}

\vskip -10 cm
\hskip 0. cm
{\Large
$t$}

\vskip 8.5 cm
\hskip 7.5 cm
{\Large
$\eta$}

\vskip -15. cm
\hskip 0 cm
{\Large
$t_m$}

\newpage
\thispagestyle{empty}
\hspace{5cm}
{\Large
\textbf{
Fig. 4b}
, Bellier-Castella et al., JCP}
\\[2cm]
\large
\begin{center}
\unitlength1mm
\begin{picture}(140,140)
\put(0,0){\makebox(140,140)[b]
{\epsfysize140mm \leavevmode \epsffile{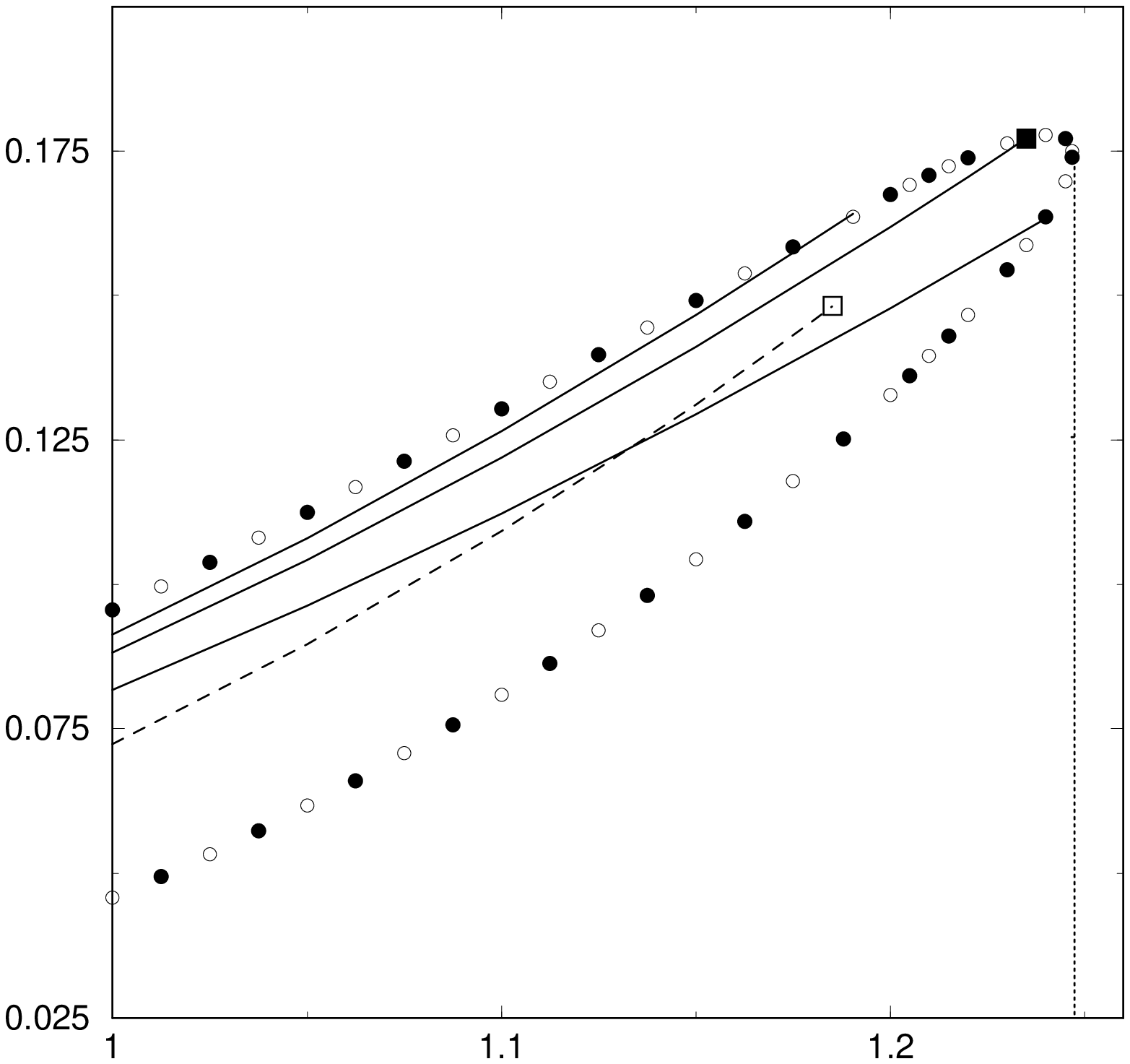} }}
\end{picture}
\end{center}

\vskip -11 cm
\hskip 0. cm
{\Large
$\overline{p}$}

\vskip 9 cm
\hskip 7.5 cm
{\Large
$t$}

\vskip -1. cm
\hskip 14. cm
{\Large
$t_m$}

\newpage
\thispagestyle{empty}
\hspace{5cm}
{\Large
\textbf{
Fig. 5a}
, Bellier-Castella et al., JCP}
\\[2cm]
\large
\begin{center}
\unitlength1mm
\begin{picture}(140,140)
\put(0,0){\makebox(140,140)[b]
{\epsfysize140mm \leavevmode \epsffile{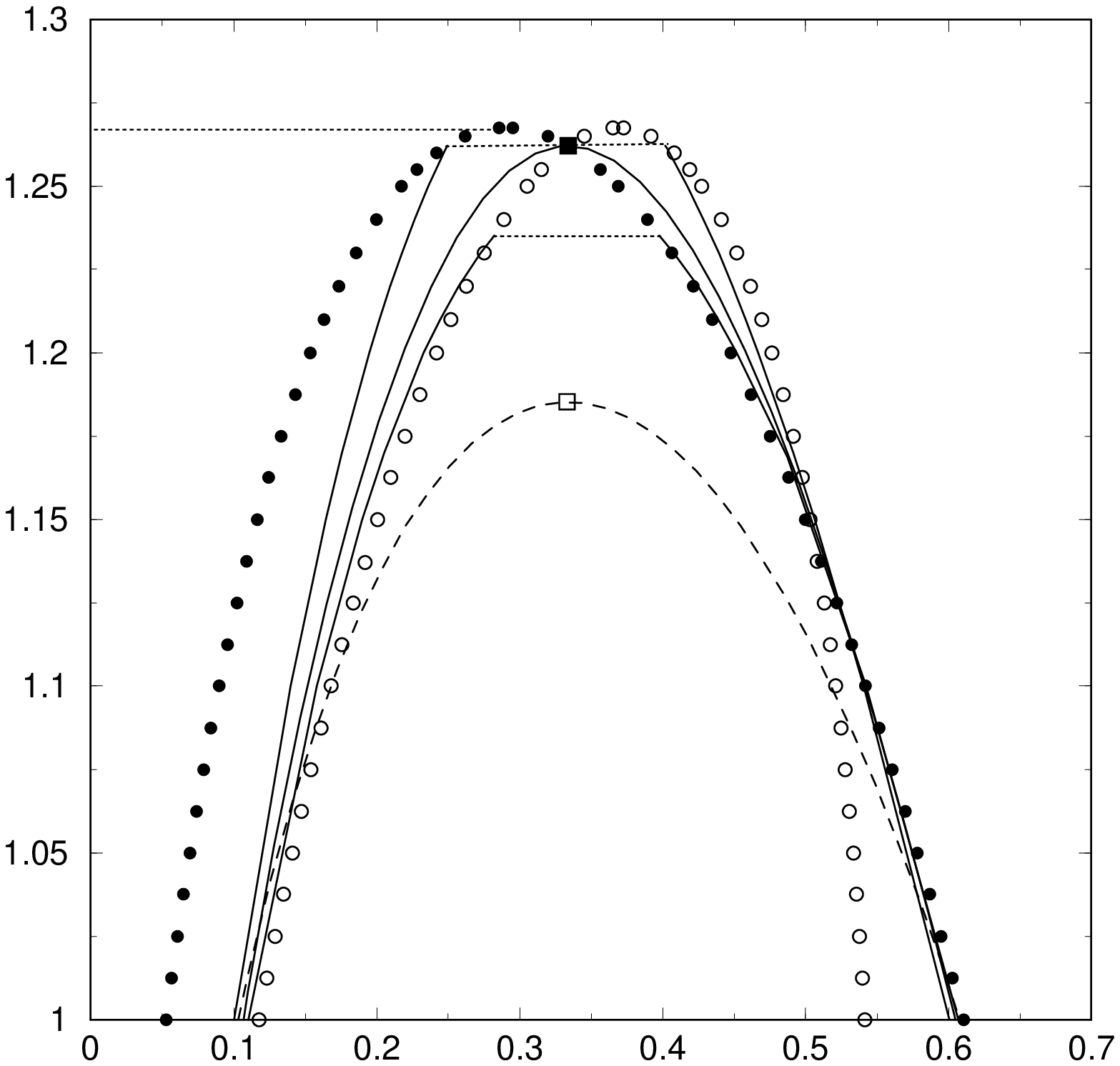} }}
\end{picture}

\vskip -9 cm
\hskip -15. cm
{\Large
$t$}

\vskip 7.5 cm
\hskip 2. cm
{\Large
$\eta$}

\vskip -13.5 cm
\hskip -14. cm
{\Large
$t_m$}

\newpage
\thispagestyle{empty}
\hspace{5cm}
{\Large
\textbf{
Fig. 5b}
, Bellier-Castella et al., JCP}
\\[2cm]
\large
\begin{center}
\unitlength1mm
\begin{picture}(140,140)
\put(0,0){\makebox(140,140)[b]
{\epsfysize140mm \leavevmode \epsffile{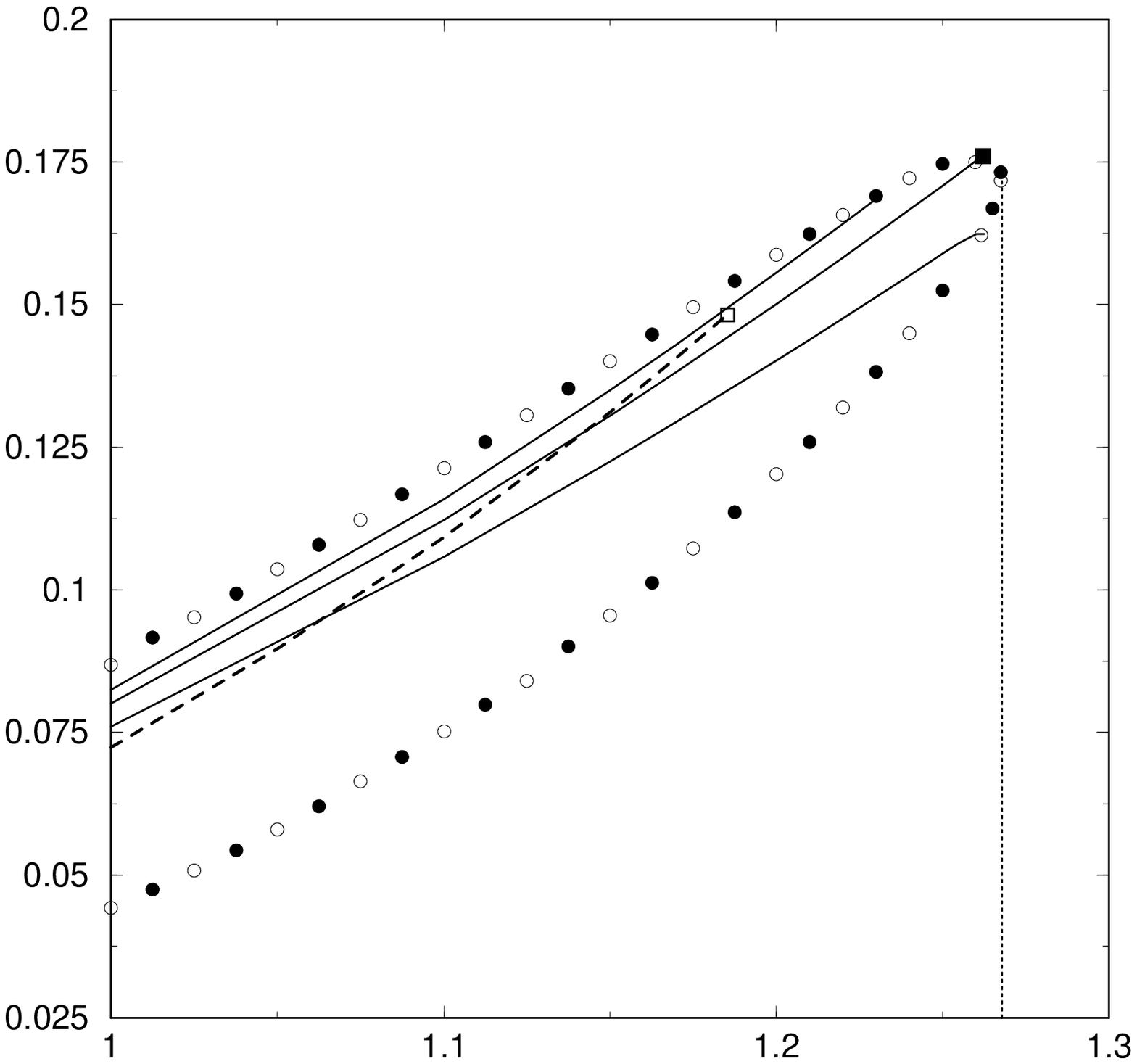} }}
\end{picture}
\end{center}

\vskip -8.5 cm
\hskip -15. cm
{\Large
$\overline{p}$}

\vskip 6.5 cm
\hskip 0 cm
{\Large
$t$}

\vskip -1. cm
\hskip 11.5 cm
{\Large
$t_m$}

\newpage
\thispagestyle{empty}
\hspace{5cm}
{\Large
\textbf{
Fig. 6}
, Bellier-Castella et al., JCP}
\\[2cm]
\large
\begin{center}
\unitlength1mm
\begin{picture}(140,140)
\put(0,0){\makebox(140,140)[b]
{\epsfysize140mm \leavevmode \epsffile{fig6.epsi} }}
\end{picture}
\end{center}

\newpage
\thispagestyle{empty}
\hspace{5cm}
{\Large
\textbf{
Fig. 7}
, Bellier-Castella et al., JCP}
\\[4cm]
\large
\begin{center}
\unitlength1mm
\begin{picture}(140,140)
\put(0,0){\makebox(140,140)[b]
{\epsfysize140mm \leavevmode \epsffile{fig7.epsi} }}
\end{picture}
\end{center}

\end{center}
\end{document}